\documentclass[a4paper]{AO4ELT}  

\usepackage{microtype}
\usepackage{biblatex}
\usepackage{amsmath,amsfonts,amssymb}
\usepackage{graphicx}
\usepackage{pst-all} 
\usepackage[colorlinks=true, allcolors=blue]{hyperref}
\usepackage[hang]{subfigure}
\usepackage{multirow}
\usepackage{aasmacros}

\makeatletter         
\def\@maketitle{
\includegraphics[width = 165mm]{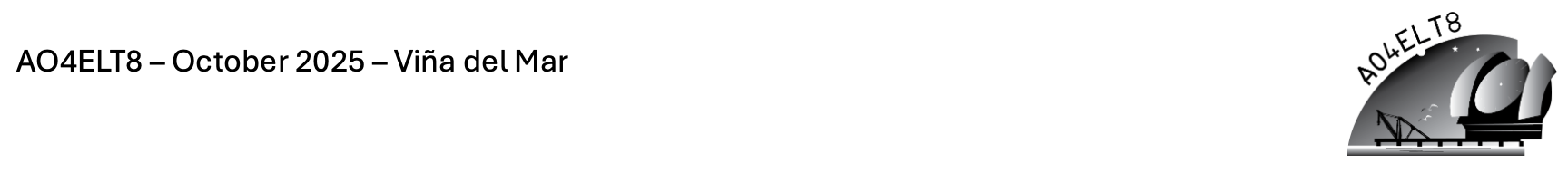}\\[8ex]
\begin{center}
{\Huge \bfseries \sffamily \@title }\\[4ex] 
{\Large  \@author}\\[4ex] 
\@date
\end{center}}

\title{MORFEO wavefront error budget}

\author[a,c]{Guido Agapito}
\author[a,c]{Lorenzo Busoni}
\author[a,c]{Cédric Plantet}
\author[a,c]{Giulia Carlà}
\author[b]{Jean-Pierre Véran}
\affil[a]{INAF Osservatorio Astrofisico di Arcetri, Largo E. Fermi 5, 50125 Firenze, Italy}
\affil[b]{Herzberg Astronomy and Astrophysics, National Research Council of Canada, 5071 West Saanich Road,
Victoria, BC V9E 2E7, Canada}
\affil[c]{ADaptive Optics National laboratory in Italy (ADONI)}

\authorinfo{Further author information: Guido Agapito: E-mail: guido.agapito@inaf.it}

\begin{document} 
\maketitle

\begin{abstract}

MORFEO (Multi-conjugate adaptive Optics Relay For ELT Observations, formerly MAORY) is the multi-conjugate adaptive optics module of the ESO Extremely Large Telescope (ELT), designed to deliver diffraction-limited performance in the near-infrared for its first-light camera MICADO. With its 12 wavefront sensors and three deformable mirrors, MORFEO stands as the largest and most complex adaptive optics system ever developed for astronomical observations.

A key aspect of its design and verification is the construction of a comprehensive wavefront error (WFE) budget, which defines the system's expected performance under a range of operating conditions. In this work, we present the structure of the MORFEO WFE budget, detailing the main contributors to the residual error and the methodology adopted to evaluate each term. The analysis includes contributions from atmospheric turbulence, optical surfaces, alignment tolerances, control residuals, and calibration uncertainties.

We also discuss the external conditions, system configurations, and key assumptions used in the derivation of the budget, highlighting the interplay between requirements and instrument design choices. Importantly, the resulting error budget is being used to support end-to-end simulations of MORFEO-assisted observations, providing essential input for assessing the scientific capabilities of the system and guiding the definition of future observing strategies.
\end{abstract}


\section{INTRODUCTION}
The Multi-conjugate Adaptive Optics Relay For ELT Observations (MORFEO\cite{2024SPIE13097E..22C}, formerly known as MAORY\cite{2021Msngr.182...13C}) is a first-light instrument for the ESO Extremely Large Telescope (ELT\cite{2024Msngr.192....3C}).
It is designed to compensate for atmospheric turbulence over a large field of view (FoV) of 60" in diameter, feeding the near-infrared camera MICADO\cite{2021Msngr.182...17D}.
The system relies on a constellation of 6 Laser Guide Stars (LGS), 3 Natural Guide Stars (NGS), and uses the ELT adaptive quaternary mirror (M4\cite{2019Msngr.178....3V}) plus two post-focal deformable mirrors (PFDMs\cite{BiasiAO4ELT8}) to achieve high Strehl Ratio (SR) and high sky coverage.

All the details of MORFEO are described in Busoni et al.\cite{BusoniAO4ELT8}, instead, this paper focus on the wavefront error budget and on the estimation of the performance.
As MORFEO approaches its final design phase, the accurate estimation of the residual Wavefront Error (WFE) becomes critical to ensure compliance with the stringent top-level requirements.
The complexity of the system—involving tomography, multiple control loops (High Order, Reference, Low Order), and interaction with the telescope structure—requires a detailed breakdown of error sources.

In previous works, we presented the baseline performance and sensitivity analyses using End-to-End (E2E) simulations\cite{2020SPIE11448E..2SA} and sky coverage estimates\cite{2022JATIS...8b1509P}.
In this work, we present the consolidated WFE budget structure.
This budget combines results from E2E simulations (PASSATA\cite{2016SPIE.9909E..7EA} and SPECULA\cite{specula2026}) and Fourier space based simulations (TipTop\cite{2020SPIE11448E..2TN}) with dedicated analytical estimates and specific sub-system analyses (e.g., thermal effects, telescope residuals, calibration errors). 

In the next sections, after describing how the budget has evolved during the different phases due to changes in the design and more accurate estimates of the different error terms in Sec.~\ref{sec:evo}, we analyze four specific observing scenarios defined in the technical specifications (Sec.~\ref{sec:scenario}), we describe the error sources (Sec.~\ref{sec:method}) and demonstrate the system's compliance (Sec.~\ref{sec:res}).

\section{EVOLUTION OF THE WAVEFRONT ERROR BUDGET}\label{sec:evo}
The estimation of MORFEO performance has evolved significantly throughout the project lifecycle, moving from statistical allocations in the early phases to a consolidated budget based on subsystem Final Design and detailed analyses.

\subsection{From Phase A to Preliminary Design}
In the early stages of the project\cite{2010aoel.confE2007D} and up to the initial Phase B, the performance estimation relied heavily on E2E simulations for the atmospheric and tomographic terms\cite{2014SPIE.9148E..6FA,2016SPIE.9909E..7BA}, while all other error sources were grouped into a single "extra" error term.
In Agapito et al. (2020)\cite{2020SPIE11448E..2SA}, this aggregate term was estimated at 105 nm RMS, serving as a placeholder for calibration errors, non-common path aberrations (NCPA), and optomechanical tolerances.

As the project approached the Preliminary Design Review (PDR) in 2022, the budget structure was expanded to explicitly identify the contributors to this extra term.
At PDR, the budget included conservative allocations for effects not simulated in the E2E tools, specifically for the ``Design, Manufacturing and Alignment of PFRO'' (Post-Focal Relay Optics) and ``Optical effects of air in the PFRO'' (internal seeing).
In Busoni et al. (2022)\cite{2022SPIE12185E..4RB}, these terms were budgeted at 100 nm RMS each, which, combined with other minor terms, led to a total "extra" error of approximately 190 nm RMS added quadratically to the simulation results.
This conservative approach was necessary to mitigate risks associated with the incomplete thermal and mechanical design at that stage.

\subsection{Towards Final Design: Refinement and Consolidation}
Entering the Final Design phase, the focus shifted towards a rigorous verification of these allocations. The successful closure of the Optical FDR in 2023 and the advanced status of the Main Structure design allowed for a reassessment of the main error contributors.
As reported in Ciliegi et al. (2024)\cite{2024SPIE13097E..22C}, a dedicated review of the error budget terms reduced the static and dynamic residuals of the PFRO.
Most notably, extensive thermal analyses and the definition of the thermal enclosure strategy led to a drastic reduction of the internal seeing contribution.

In the current consolidated budget presented in this work, the ``extra'' terms have been fully ``exploded'' into specific line items derived from analytical models or sub-system specifications.
Comparing the current values with the PDR baseline:
\begin{itemize}
    \item The \textbf{Internal Seeing} (Optical effects of air) has been reduced from 100 nm to \textbf{20 nm RMS}, supported by high-fidelity CFD simulations of the thermal control system.
    \item The \textbf{PFRO Residuals} (Design, Manufacturing, and Alignment) have been refined from 100 nm to \textbf{70 nm RMS}, based on the as-built specifications for the optical manufacturing and estimation of the compensation provided by day-time calibration and truth sensing.
    \item A specific allocation for \textbf{Vibrations} of the post-focal DMs ($\sim$35 nm) and \textbf{Calibration Errors} ($\sim$32 nm) has been introduced, replacing generic margins.
\end{itemize}
This evolution marks the transition from a ``top-down'' requirement allocation to a ``bottom-up'' performance verification.
While the PDR budget contained significant margins to cover design uncertainties, the current budget reflects the expected physical behavior of the hardware, with a specific contingency term (50 nm) kept separate to manage residual risks during the Assembly, Integration, and Verification (AIV) phase.

\section{SCENARIOS AND REQUIREMENTS}\label{sec:scenario}
To validate the design, we evaluate the WFE budget against four specific requirements defined in the MORFEO Technical Specification. These scenarios cover different atmospheric conditions and performance metrics (Strehl Ratio, Ensquared Energy, and FWHM).

\begin{table}[ht]
\caption{Summary of the technical specification requirements and atmospheric conditions\cite{2013aoel.confE..89S}. The Ensquared Energy (EE) is computed in a $16 \times 16$ mas aperture. The NGS asterisms are described in Sec.~\ref{sec:scenario}.}
\label{tab:requirements}
\begin{center}
\begin{tabular}{|c|c|c|c|c|c|c|}
\hline
\rule[-1ex]{0pt}{3.5ex} \textbf{Case ID} & \textbf{Atmosp.} & \textbf{Wavel. [nm]} & \textbf{SR}  & \rule[-1ex]{0pt}{3.5ex} \textbf{FWHM [mas]}  & \textbf{EE [\%]} & \textbf{FoV $\phi$ ["]} \\
\hline
\rule[-1ex]{0pt}{3.5ex}  \multirow{3}{*}{R-MAO-80} & \multirow{3}{*}{Q1 (Best)} & 2200  & 0.60 & 13 & 38 & \multirow{3}{*}{20} \\
\rule[-1ex]{0pt}{3.5ex} {} & {} & 1250 & 0.20 & 9 & 19 & {} \\
\rule[-1ex]{0pt}{3.5ex} {} & {} & 850 & 0.05 & 8 & 6 & {} \\
\hline
\rule[-1ex]{0pt}{3.5ex} \multirow{2}{*}{R-MAO-82} & \multirow{2}{*}{Median} & 2200 & 0.44 & 13.5 & 28 & \multirow{2}{*}{60} \\
\rule[-1ex]{0pt}{3.5ex} {} & {} & 1250 & 0.08 & 9 & 10 & {} \\
\hline
\rule[-1ex]{0pt}{3.5ex} R-MAO-83 & Q4 (Poor) & 2200 & 0.25 & 14 & 16 & 20 \\
\hline
\rule[-1ex]{0pt}{3.5ex} R-MAO-168 & Q1 (1 NGS) & 2200 & 0.50 & 14 & 34 & 20 \\
\hline
\end{tabular}
\end{center}
\end{table}

Tab. \ref{tab:requirements} summarizes these cases. R-MAO-82 represents the nominal operation in median seeing conditions ($\sim 0.65"$ at 500nm, see Ref.~\cite{2013aoel.confE..89S}), requiring uniform correction over the full MICADO FoV.
The other cases focus on peak performance (Q1 conditions) or robustness (Q4 conditions).
For each case, specific NGS asterisms were selected to represent realistic sky coverage conditions, over 50\% of the south galactic poles.
These are:
\begin{itemize}
    \item R-MAO-80: an NGS configuration with at least two stars, where the brightest is as faint as H=16, and the other as faint as H=20 and R=21.5 – with the geometrical barycenter located within 40” of the MICADO FoV center.
    \item R-MAO-82 and 83: an NGS configuration with at least two stars, where the brightest is as faint as H=17.5, and the other as faint as H=20 and R=21.5 – with the geometrical barycenter located within 40” of the MICADO FoV center.
    \item R-MAO-168: one NGS as faint as H=19 and R=21, and located within 65'' of the MICADO FoV center.
\end{itemize}

\section{METHODOLOGY AND ERROR SOURCES}\label{sec:method}
The WFE budget is constructed by categorizing error sources into main groups.
A significant portion of the WFE is derived directly from time-domain E2E simulations, which include atmospheric turbulence, DM fitting, WFS noise, and control loop latency.
However, several effects cannot be efficiently simulated in a standard E2E run (due to computational load or timescale differences) and are estimated via separate analyses and added quadratically.

Please note that several trade-off analyses were performed in order to select a system design that corresponds to the WFE budget\cite{2022JATIS...8b1514F,2022JATIS...8b1505A,2022SPIE12185E..4PA,2024SPIE13097E..54M,2024SPIE13096E..5LD,2024SPIE13096E..5KM}.

The global WFE $\sigma_{tot}$ is defined as:
\begin{equation}
    \sigma_{tot}^2 = \sigma_{HO}^2 + \sigma_{LO}^2 + \sigma_{Focus}^2 + \sigma_{Ref}^2 + \sigma_{Relay}^2 + \sigma_{Calib}^2 + \sigma_{Tel}^2 + \sigma_{Other}^2
\end{equation}

In the following sections, we detail the contributors to these terms.

\subsection{High Orders (HO)}
This term encompasses errors related to LGS tomography and high-order correction (excluding tip-tilt and focus).
\begin{itemize}
    \item \textbf{Tomographic error:} This is the dominant term (approx. 135 nm in median conditions). It arises from the finite number of LGS and the vertical distribution of turbulence. 
    \item \textbf{Generalized Fitting:} Error due to the limited number of DMs and actuators. This error depends on the size of the optimized FoV.
    \item \textbf{Measurement Noise \& Temporal Error:} Derived from E2E simulations considering the LGS flux (approx. 600 ph/subap/frame) and the loop delay (integrator gain optimization).
\end{itemize}

\subsection{Low Orders and Focus}
These terms are derived from the performance of the Natural Guide Star (NGS) loops (Low Order WFS in the infrared):
\begin{itemize}
    \item \textbf{Windshake:} Residual tip-tilt jitter caused by wind load on the telescope (mainly M2). This is a major contributor, ranging from 63 nm (Q1) to 191 nm (Q4).
    \item \textbf{Tomography \& Noise:} Tip-Tilt decorrelation of turbulence at the NGS locations and measurement noise.
    \item \textbf{Field Average Focus:} MORFEO controls focus using NGS (see Ref.~\cite{AgapitoAO4ELT8} for more details on focus control on MORFEO).
    This term includes anisoplanatism and temporal errors.
\end{itemize}
While this section addresses the WFE contribution of low-order modes, the specific impact on pointing stability and the detailed platescale budget
have been analyzed using the approach described by Carlà et al. 2022~\cite{2022SPIE12185E..0OC}

\subsection{Reference Loop}
The Reference loop uses 3 visible NGS WFS to correct for low-order modes that are poorly sensed by LGS (e.g., sodium layer altitude variations) or affected by non-common path aberrations (NCPA).
\begin{itemize}
    \item \textbf{Correction Residual:} This is a negative term in the budget. The main E2E simulation includes truncation errors due to LGS spot elongation. The reference loop corrects these slowly evolving aberrations by effectively ``subtracting'' this error from the raw E2E output. However, this value changes when only one NGS is available (it is scaled by a factor 0.6), as the reference loop cannot correct for aberrations at different altitudes to the ground layer in this case.
    \item \textbf{Sodium Profile Variations:} Variations in the sodium layer profile create quasi-static aberrations. We budget 15 nm rms for residual errors after Reference loop correction.
    \item \textbf{Propagation of Atmospheric Disturbance:} The Reference loop has a low bandwidth but can inadvertently propagate some atmospheric turbulence into the main loop. This is estimated at $\sim 30$ nm for median conditions.
\end{itemize}

\subsection{MORFEO Relay and Internal Errors}
This category includes errors intrinsic to the instrument hardware:
\begin{itemize}
    \item \textbf{Optics Quality:} Residual design, manufacturing, and alignment errors of the Post Focal Relay Optics (PFRO) are estimated at 70 nm rms (the analysis of this error is presented in Ref.~\cite{2024SPIE13097E..4RP}). This includes thermo-elastic deformations.
    For RMAO-168, this error introduces an additional contribution of approximately 30 nm rms, increasing the error term to 76 nm. This happens because, when only one NGS is available, the tomographic truth-sensing correction is unavailable, and we must rely solely on look-up tables, which cannot track the slow, field-dependent thermal drifts of the NCPA over the observation night.    
    \item \textbf{Internal Seeing:} Optical effects of air inside the MORFEO enclosure. Based on thermal analysis, we budget 20 nm rms.
    \item \textbf{DM Fitting \& Vibrations:} The real influence functions of the PFDMs differ from ideal ones (30 nm fitting error per DM). Additionally, interface motion amplification can induce high-frequency vibrations on the PFDMs ($\sim 35$ nm rms).
\end{itemize}

\subsection{Telescope and Interface}
Errors originating from the ELT structure:
\begin{itemize}
    \item \textbf{Telescope Residuals:} Derived from the ESO telescope wavefront perturbations data package (M1/M2 residuals).
    \item \textbf{Low Wind Effect (LWE):} Caused by temperature differentials across the telescope spiders. In Q1 conditions, where LWE is most prominent, we budget 50 nm rms, assuming correction by the MORFEO LIFT sensor implemented in the Low Order WFS\cite{2010OptL...35.3036M,2022SPIE12185E..56A,PlantetAO4ELT8}.
    \item \textbf{Off-axis Aberrations:} Telescope aberrations vary across the field. We assume the MCAO system corrects 50\% of these, leaving a residual of $\sim 24$ nm for the 60" FoV.
\end{itemize}

\section{RESULTS AND PERFORMANCE}\label{sec:res}
\begin{figure}[ht]
    \begin{center}
        \subfigure[K band SR.\label{fig:SRK}]
        {\includegraphics[width=0.49\columnwidth]{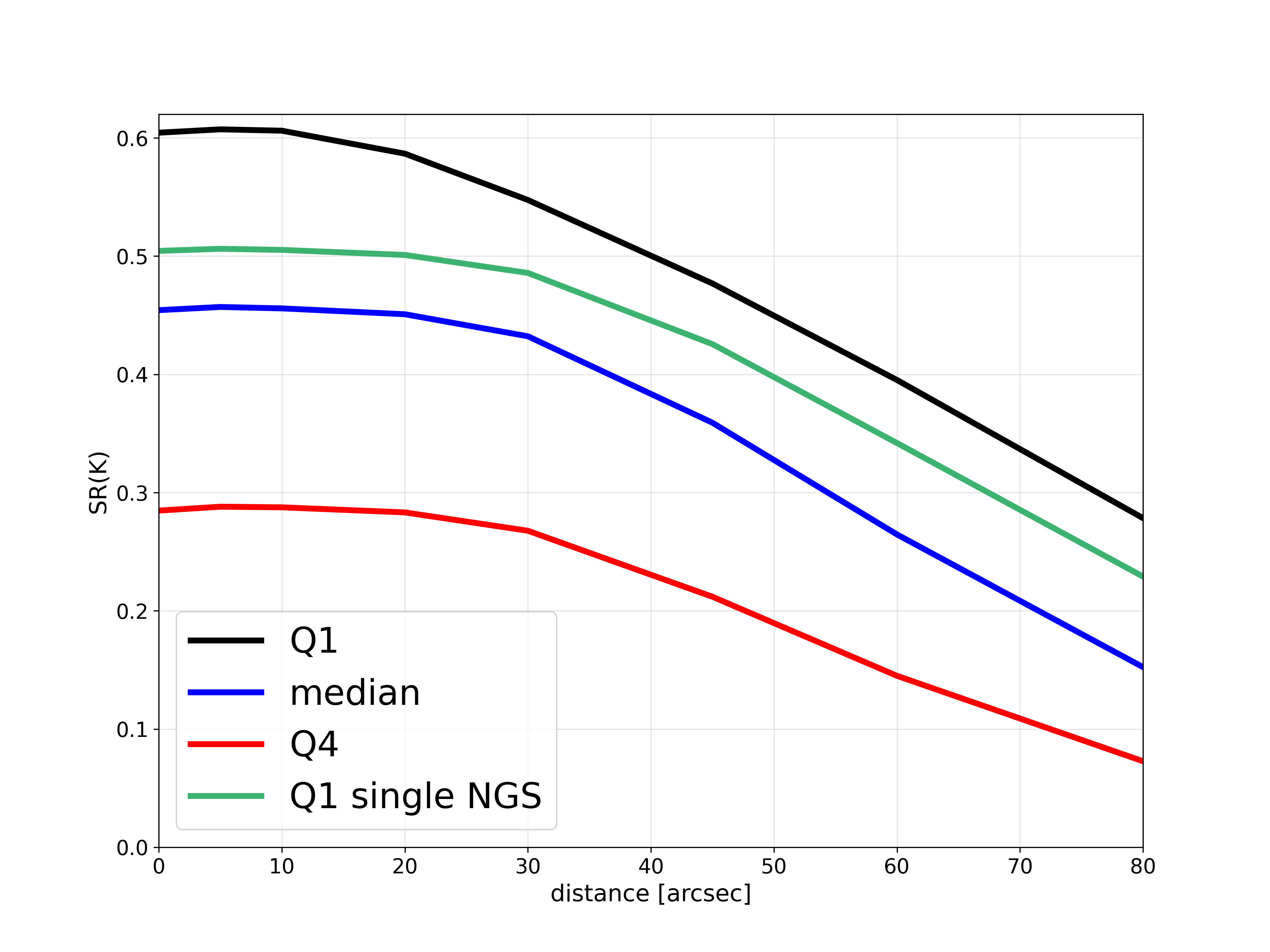}}
        \subfigure[SR in Q1 atmospheric conditions. \label{fig:SRQ1}]
        {\includegraphics[width=0.49\columnwidth]{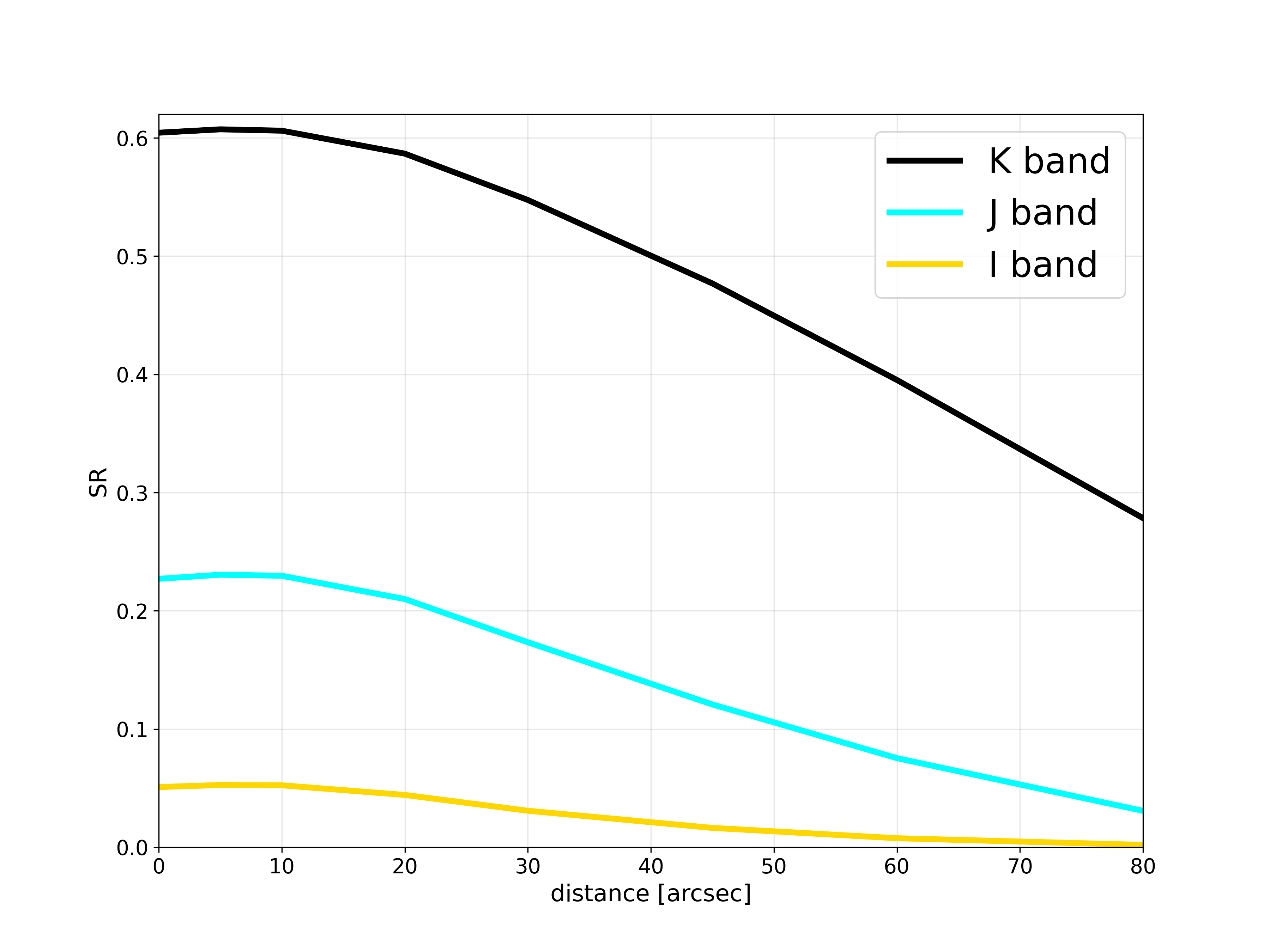}}
    \end{center}
    \caption{Summary of the SR as a function of the off-axis angle of MORFEO for different wavelengths and atmospheric conditions considering a zenith angle of 30 deg. The NGS asterisms are described in Sec.~\ref{sec:scenario}.}
    \label{fig:SR}
\end{figure}
\begin{figure}[ht]
    \begin{center}
        \subfigure[K band FWHM. \label{fig:FWHMK}]
        {\includegraphics[width=0.49\columnwidth]{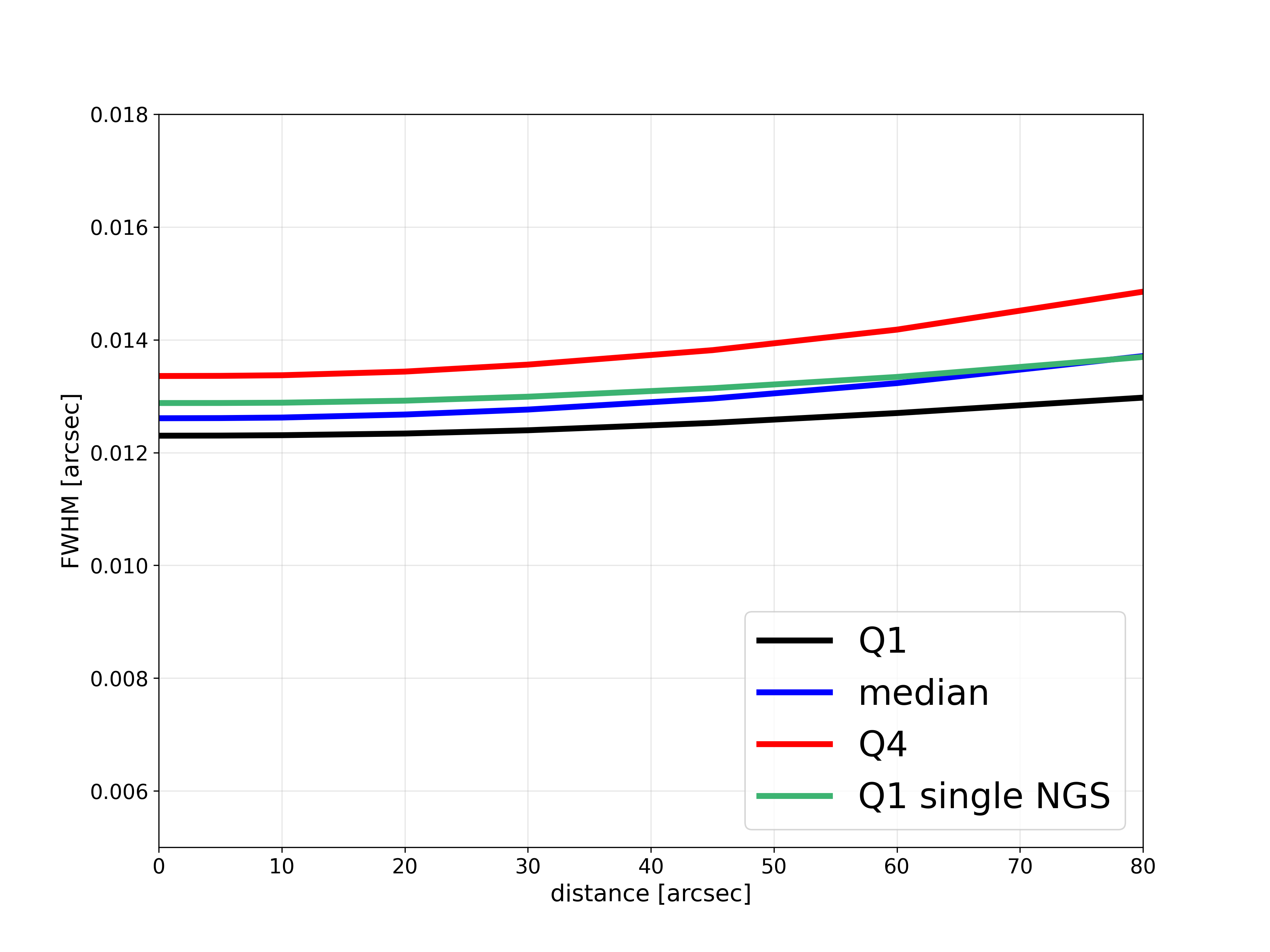}}
        \subfigure[FWHM in Q1 atmospheric conditions. \label{fig:FWHMQ1}]
        {\includegraphics[width=0.49\columnwidth]{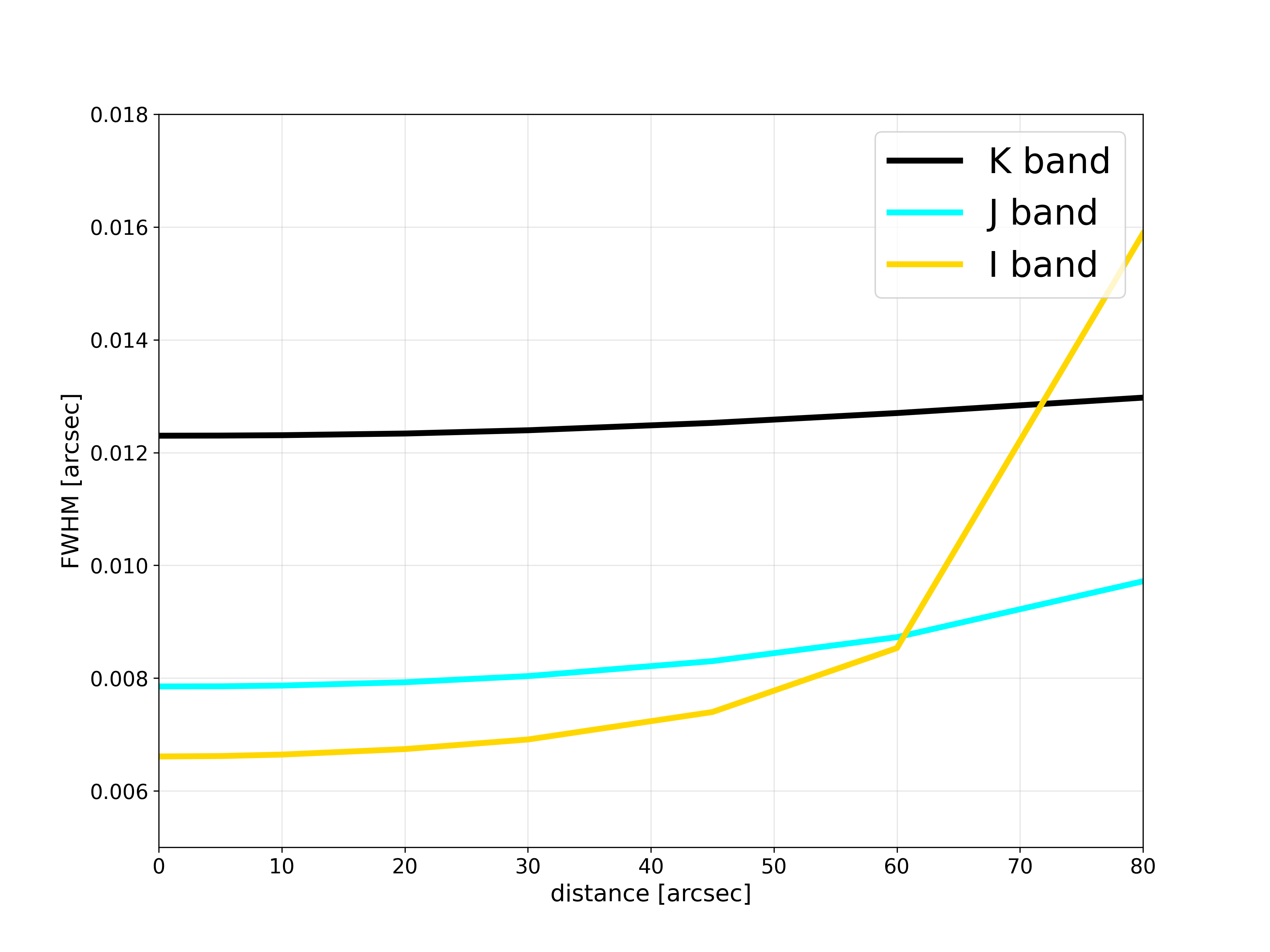}}
    \end{center}
    \caption{Summary of the FWHM as a function of the off-axis angle of MORFEO for different wavelengths and atmospheric conditions considering a zenith angle of 30 deg. The NGS asterisms are described in Sec.~\ref{sec:scenario}.}
    \label{fig:FWHM}
\end{figure}
\begin{figure}[ht]
    \begin{center}
        \subfigure[K band Ensquared Energy d=16mas. \label{fig:EEK}]
        {\includegraphics[width=0.49\columnwidth]{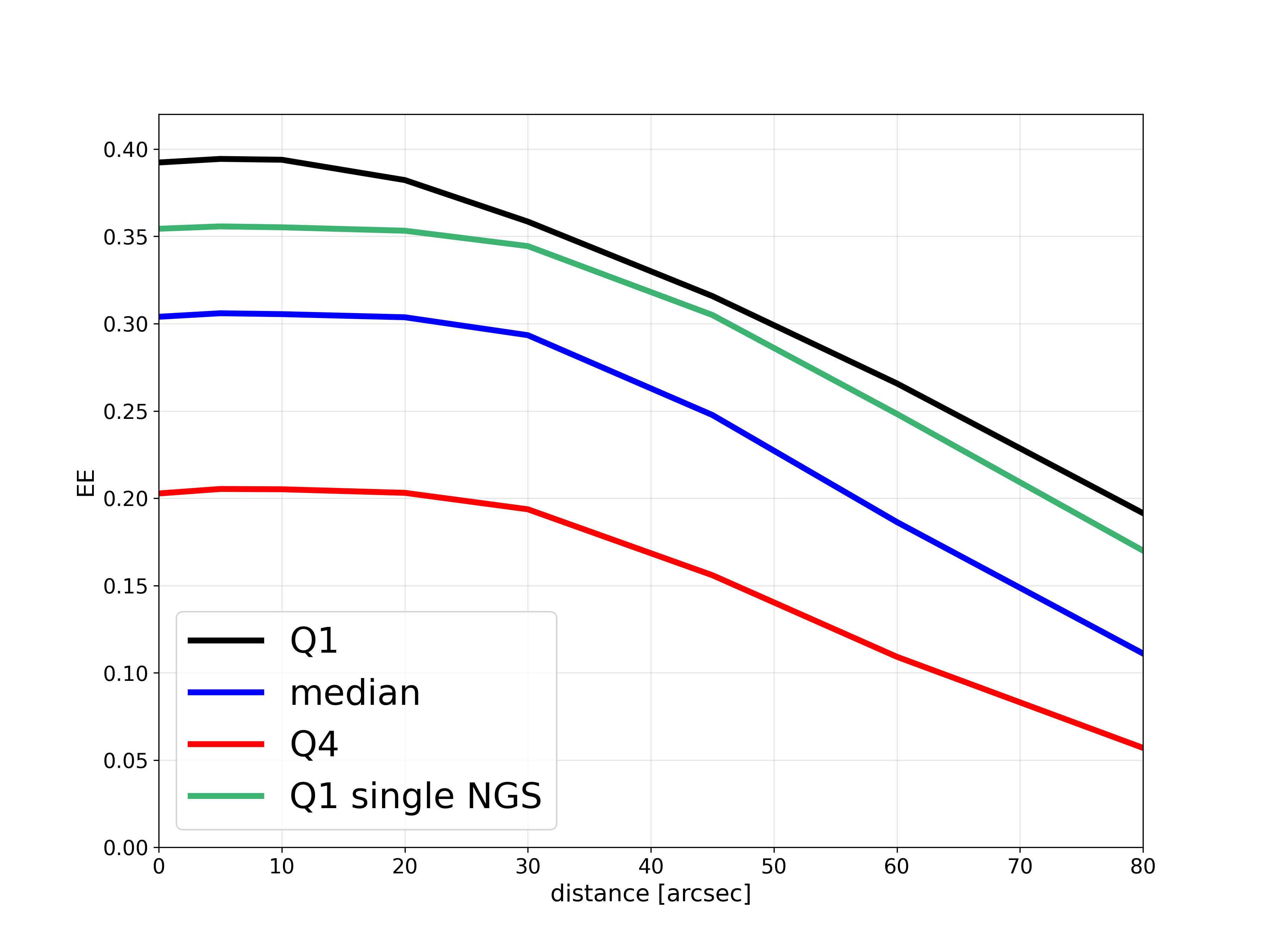}}
        \subfigure[Ensquare Energy d=16mas in Q1 atmospheric conditions. \label{fig:EEQ1}]
        {\includegraphics[width=0.49\columnwidth]{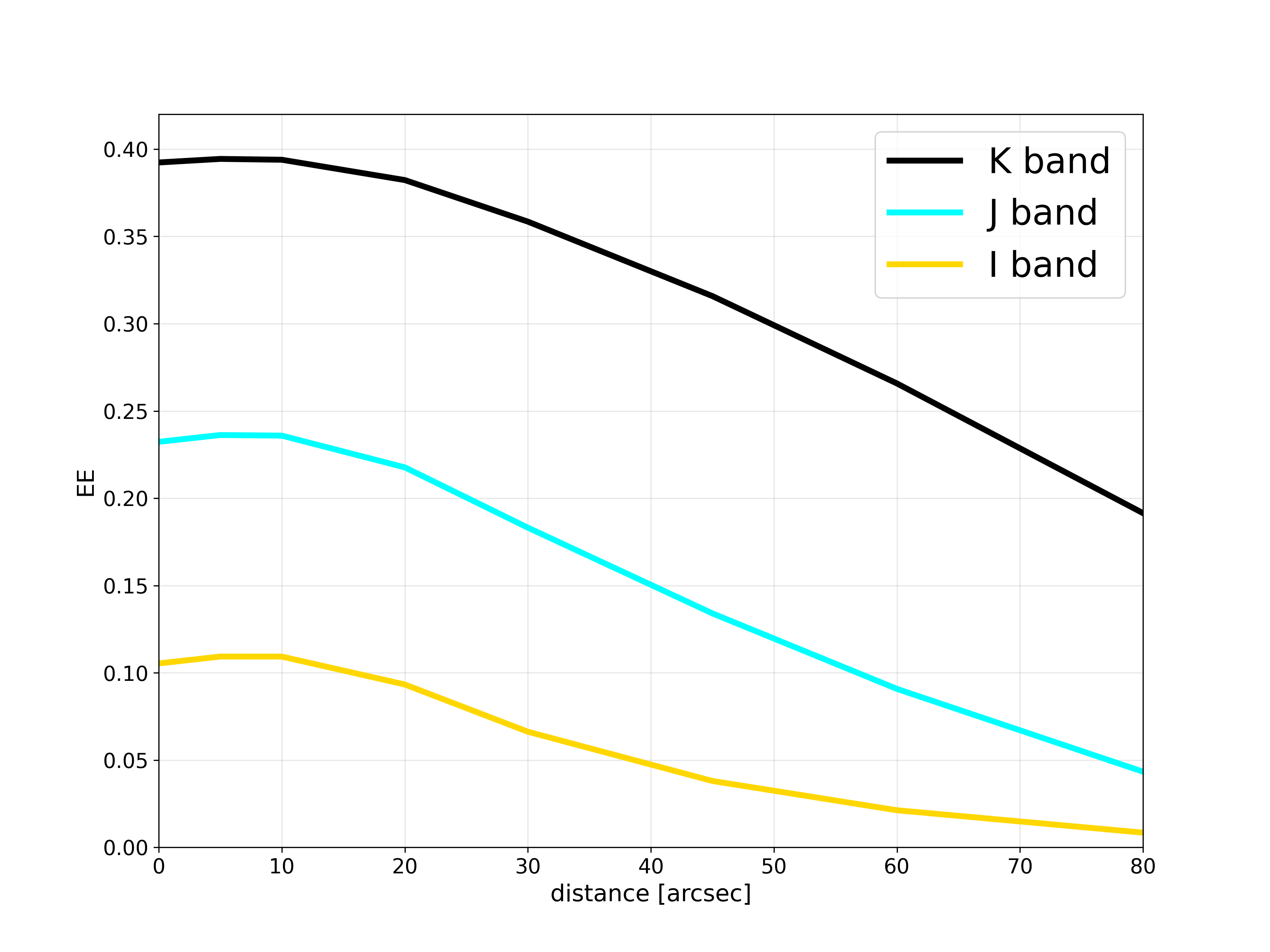}}
    \end{center}
    \caption{Summary of the Ensquare Energy d=16mas as a function of the off-axis angle of MORFEO for different wavelengths and atmospheric conditions considering a zenith angle of 30 deg. The NGS asterisms are described in Sec.~\ref{sec:scenario}.}
    \label{fig:EE}
\end{figure}
We combined all error sources to produce the final budget for the four reference cases.
The results are summarized in Tab. \ref{tab:budget_summary} and \ref{tab:performance_summary} and in Fig. \ref{fig:SR}, \ref{fig:FWHM} and  \ref{fig:EE}.
The SR, FWHM and EE are estimated from PSFs as described in the Appendix~\ref{sec:appendix_metrics}.

\begin{table}[ht]
\caption{Wavefront Error Budget Summary for the four requirement scenarios.}
\label{tab:budget_summary}
\begin{center}
\begin{tabular}{|l|c|c|c|c|}
\hline
\textbf{Term [nm rms]} & \textbf{R-MAO-82} & \textbf{R-MAO-80} & \textbf{R-MAO-83} & \textbf{R-MAO-168} \\
\hline
\rule[-1ex]{0pt}{3.5ex} High Orders & 215 & 164 & 271 & 164 \\
\rule[-1ex]{0pt}{3.5ex} Low Orders (TT) & 140 & 96 & 274 & 153 \\
\rule[-1ex]{0pt}{3.5ex} Field Focus & 116 & 52 & 179 & 92 \\
\rule[-1ex]{0pt}{3.5ex} Reference Loop & 56 & 37 & 53 & 57 \\
\rule[-1ex]{0pt}{3.5ex} MORFEO Relay & 91 & 91 & 91 & 96 \\
\rule[-1ex]{0pt}{3.5ex} Calibration & 32 & 31 & 33 & 31 \\
\rule[-1ex]{0pt}{3.5ex} Telescope & 55 & 71 & 50 & 71 \\
\rule[-1ex]{0pt}{3.5ex} Atm. Chromatism & 24 & 17 & 36 & 17 \\
\rule[-1ex]{0pt}{3.5ex} Rayleigh Scattering & 36 & 20 & 20 & 20 \\
\rule[-1ex]{0pt}{3.5ex} Others & 50 & 50 & 50 & 50 \\
\rule[-1ex]{0pt}{3.5ex} Contingency & 50 & 50 & 50 & 50 \\
\hline
\textbf{Total WFE [nm]} & \textbf{318} & \textbf{245} & \textbf{449} & \textbf{288} \\
\hline
\end{tabular}
\end{center}
\end{table}

\begin{table}[ht]
\caption{Performance summary estimated for the four reference scenarios. The Ensquared Energy (EE) is computed in a $16 \times 16$ mas aperture. The NGS asterisms are described in Sec.~\ref{sec:scenario}.}
\label{tab:performance_summary}
\begin{center}
\begin{tabular}{|l|l|c|c|c|c|}
\hline
\rule[-1ex]{0pt}{3.5ex} \textbf{Case ID} & \textbf{Conditions} & \textbf{Band} & \textbf{SR} & \textbf{FWHM [mas]} & \textbf{EE [\%]} \\
\hline
\rule[-1ex]{0pt}{3.5ex} \multirow{3}{*}{R-MAO-80} & \multirow{3}{*}{Q1 atmo. FoV: 20"} & K (2.2 $\mu$m) & 0.606 & 12.3 & 39.4 \\
\rule[-1ex]{0pt}{3.5ex}  & & J (1.25 $\mu$m) & 0.230 & 7.9 & 23.6 \\
\rule[-1ex]{0pt}{3.5ex}  & & I (0.85 $\mu$m) & 0.052 & 6.6 & 10.9 \\
\hline
\rule[-1ex]{0pt}{3.5ex} \multirow{2}{*}{R-MAO-82} & \multirow{2}{*}{Median atmo. FoV: 60"} & K (2.2 $\mu$m) & 0.445 & 12.7 & 30.1 \\
\rule[-1ex]{0pt}{3.5ex}  & & J (1.25 $\mu$m) & 0.103 & 8.7 & 12.6 \\
\hline
\rule[-1ex]{0pt}{3.5ex} R-MAO-83 & Q4 atmo.FoV: 20" & K (2.2 $\mu$m) & 0.288 & 13.4 & 20.5 \\
\hline
\rule[-1ex]{0pt}{3.5ex} R-MAO-168 & Q1 atmo. FoV: 20" & K (2.2 $\mu$m) & 0.506 & 12.9 & 34.8 \\
\hline
\end{tabular}
\end{center}
\end{table}

In the median condition (R-MAO-82), the total WFE is 317 nm, yielding a K-band Strehl Ratio of 0.44 (using the Maréchal approximation), which meets the requirement.
The ``Best'' condition (R-MAO-80) achieves 245 nm WFE (SR 0.61).
The ``Poor'' condition (R-MAO-83) is dominated by windshake and generalized fitting error due to strong turbulence (seeing at zenith 1.04"), resulting in 448 nm WFE.
Please note that while the estimated SR using the Maréchal's approximation, 0.19, is slightly below the requirement, the PSF from E2E simulation shows a higher SR value, above the requirement. The discrepancy is expected because the small WFE regime of Maréchal's formula fails in this highly turbulent scenario.

\section{CONCLUSION}
We have presented the detailed Wavefront Error Budget for MORFEO. The budget is built upon a hybrid approach combining high-fidelity E2E simulations for dynamic atmospheric terms and analytical estimates for static and instrumental terms. The analysis confirms that the MORFEO design is capable of meeting the ambitious performance requirements of the ELT, delivering high Strehl Ratios in the near-infrared. This budget serves as a living document to guide the assembly, integration, and verification phases of the instrument.

\acknowledgments
This work was funded by the MORFEO consortium.

\appendix

\section{METHODOLOGY FOR SCIENTIFIC METRIC ESTIMATION}
\label{sec:appendix_metrics}

While the SR can be approximated from the total WFE variance using the Maréchal approximation, scientific merit functions such as the FWHM and EE require a direct computation from the Point Spread Function (PSF).
Since it is non-trivial to analytically estimate the impact of error budget terms not included in the E2E simulations on these specific metrics, we adopted a hybrid post-processing approach.

The estimation procedure consists of the following steps:

\begin{enumerate}
    \item \textbf{Extraction of E2E Residuals:} We extract the instantaneous wavefront error maps from the PASSATA / SPECULA E2E simulations.The dataset covers 4500 iterations (corresponding to 9 seconds of physical time) computed along 43 directions distributed across the scientific Field of View (FoV): one on-axis direction and 6 rings of uniformly distributed directions at radii of 5, 10, 20, 30, 45, 60, and 80".

    \item \textbf{Injection of Missing Error Terms:} To account for the error budget terms described in Section 4 (e.g., calibration errors, static telescope aberrations, vibrations) which are not physically modeled in the E2E loop, we generate random wavefront error maps for each step. These maps are defined to have an RMS equal to the value of the terms not included in the simulation.
    
    Based on the nature of these missing terms (which are dominated by residual windshake, telescope drifts, and low-order calibration residuals), we assume a Power Spectral Density (PSD) dominated by low spatial frequencies. Specifically, we allocate 90\% of the RMS to Tip-Tilt, while the remaining high-order RMS is distributed such that it scales with the reciprocal of the mode index.

    \item \textbf{PSF Computation:} For each simulation step, the ``extra'' random wavefront is summed with the instantaneous E2E residual. We then compute the monochromatic Short Exposure PSF with a sampling of $0.143 \, \lambda/D$ (where $\lambda$ is 850, 1250, or 2200 nm, and $D=38.5$ m).

    \item \textbf{Metric Extraction:} The Short Exposure PSFs are averaged over time to produce the Long Exposure PSF. From this final PSF, the SR, EE (computed in a $16 \times 16$ mas square aperture), and FWHM (derived from the profile) are calculated.

    \item \textbf{Field Averaging:} Finally, the metrics are averaged over the FoV of interest (e.g., radius $R=10"$ or $R=30"$ depending on the specific requirement). The average is weighted by the area $A(i)$ corresponding to each direction $i$:
    \begin{equation}
        \text{Metric}_{ave} = \frac{\sum_{i=0}^{42} \text{Metric}(i) A(i)}{\sum_{i=0}^{42} A(i)}
    \end{equation}
    The final values presented in this work are the average of the results obtained from 5 independent simulation seeds.
\end{enumerate}

\printbibliography 

\end{document}